# Machine Learning-based Optimal Control for Colloidal Self-Assembly

Andres Lizano-Villalobos[1]‡, Fangyuan Ma[2]‡, Wentao Tang[3], Wei Sun[1,2], Xun Tang[1]*

[1]Cain Department of Chemical Engineering, Louisiana State University, Baton Rouge, LA 70803, USA

[2]College of Chemical Engineering, Beijing University of Chemical Technology, Beijing 100029, China

[3]Department of Chemical and Biomolecular Engineering, North Carolina State University, Raleigh, NC 27695, USA

**KEYWORDS:** Machine Learning, Optimal Control, Deep Q-Learning, Graph Convolutional Neural Network, Colloidal Self-Assembly

## ABSTRACT

Achieving precise control of colloidal self-assembly into specific patterns remains a longstanding challenge due to the complex process dynamics. Recently, machine learning-based state representation and reinforcement learning-based control strategies have started to accumulate popularity in the field, showing great potential in achieving an automatable and generalizable approach to producing patterned colloidal assembly. In this work, we adopted a machine learning-based optimal control framework, combining unsupervised learning and graph convolutional neural work for state observation with deep reinforcement learning-based optimal control policy calculation, to provide a data-driven control approach that can potentially be generalized to other many-body self-assembly systems. With Brownian dynamics simulations, we demonstrated its superior performance as compared to traditional order parameter-based state description, and its efficacy in obtaining ordered 2-dimensional spherical colloidal self-assembly in an electric field-mediated system with an actual success rate of 97%.

## MAIN TEXT

Controlling colloidal self-assembly, the spontaneous formation of ordered structures of colloidal particles, [1] to engineer novel structures holds great potential for making functional materials with unique mechanical, photonic, electrical, and magnetic properties necessary for applications in photonic crystals, energy-saving displays, or high-efficiency solar cells. [2-5] To manufacture specific colloidal assemblies, a wide range of approaches have been developed, spanning from designing particle properties such as particle surface properties and geometry, [6-8] to substrate-guided self-assembly, [9, 10] and to external force-mediated assembly. [11-13] Among those, manipulating colloidal self-assembly with external fields, such as electric, magnetic, acoustic and optical fields, holds unique advantages in terms of speed, ease of implementation, and operation design flexibility such as frequency and strength, that has acclaimed vast popularity. [14-18]

Traditional external field-mediated colloidal self-assembly typically leverages equilibrium assembly for thermodynamically stable configurations, where the system settings such as the media concentration, magnitude and frequency of the applied fields, are fixed and the assembly is driven by local interactions among the particles. [19] Such an equilibrium assembly process could suffer from local energy minima that the assembly can be trapped in metastable states, forming disordered structures that are time consuming to address. [20] Alternatively, out-of-equilibrium external field-mediated colloidal self-assembly leverages the flexibility of manipulating the external field settings during an assembly process to dynamically maneuver the process dynamics, offering a rapid assembly approach. [21] Despite some experimental successes, [22-27] controlling colloidal assembly into desired configurations with external fields remains a significant challenge, as external interventions also tend to introduce defects, compromising the quality of the assembly. [13, 27, 28] To address this assembly quality-time efficiency problem, optimal feedback control of colloidal self-assembly using out-of-equilibrium manipulation with external forces has received increasing interests, given the capability of using feedback to inform the operation and to ensure the assembly quality. [13, 24, 25, 28-33]

Optimal control for colloidal self-assembly has been pursued with model predictive control, [29] dynamic programming, [27, 30, 31] and more recently reinforcement learning. [25, 34] Two foundations to an effective feedback control include (i) accurate state (i.e. assembly configuration in this study) description used as the feedback, and (ii) a reliable understanding of the system dynamics in response to control actions. The state-of-the-art colloidal self-assembly state description is achieved with order parameters, which are mathematical equations defined to capture the properties of the target configurations, such as the radius of gyration, $R_g$, [35, 36] local ordering $C_6$, the average number of nearest neighbors for a hexagonal assembly, [37, 38] and the global ordering $\psi_6$, which captures the particle-particle bond orientation. [37-39] Understanding the system dynamics is typically achieved with modeling, such as Brownian Dynamics simulation, [40, 41] and Markov State Models [42, 43] etc., which are developed either based on first principles or data collected from the physical system. Despite their popularity, order parameters are hard to identify, and they are problem specific with limited applicability to other systems and problems, whereas probing the system dynamics with models would always suffer from modeling error.

Over the past several years, machine learning has emerged as a promising solution to the aforementioned challenges in colloidal assembly state description and process dynamics learning. [44] For example, in [45, 46], the authors deployed convolutional neural network and unsupervised learning to analyze and classify 2-dimensional spherical colloidal self-assembly images into different states, and in [47], the authors used cloud data point analysis to quantify the configuration of 3-dimensional colloidal self-assembly. In [48], the authors used graph neural network to investigate the potential energy in a colloidal self-assembly system, based on stochastic particle movement trajectories. In [25], Zhang et. al applied reinforcement learning with order parameters to assembly eclipse colloidal structures in an electric field, and in [34], the authors presented a tutorial study on how to implement reinforcement learning for colloidal self-assembly. Other notable examples include [33, 34, 49-62]. Despite its great prospect, a complete machine learning-based colloidal self-assembly optimal control framework, covering state description and control strategy design is yet to be developed.

In this work, we present a complete machine learning-based optimal control framework for colloidal self-assembly as in Figure 1, which also holds potential for applications on other many-body self-assembly systems. Our framework features a graph convolutional neural network (GCN)-based state description, unsupervised clustering-enabled state classification, and a deep Q-learning (DQN) algorithm for optimal control policy optimization. To demonstrate the efficacy of the proposed framework, we utilize an experimentally validated Brownian Dynamics (BD) simulation model as a surrogate system for a quadrupole AC electric field-mediated colloidal self-assembly experimental system. [27]

The colloidal system contains 300 spherical $SiO_2$ colloidal particles, suspended in a batch container, which is surrounded by four evenly spaced electrodes as in Figure 1. The particle movement is governed by Brownian dynamics, and can be maneuvered by manipulating the voltage level of the applied electric field, with its frequency fixed to generate a compressing force to maintain the assembly at the center of the container. The system was previously studied with in [27], using order parameters $C_6$ and $\psi_6$ as the state descriptors, and dynamic programing for offline optimal policy calculation to guide the selection of the control action, composed by four discretized voltage levels, represented by dimensionless metric $\lambda = 0.2, 0.9, 2.0, 19.7$, with the higher $\lambda$ value corresponding to a stronger compressing force. Given the proven feasibility of achieving the ordered hexagonal colloidal assembly in the system, this quadrupole electric field-mediated system constructs an ideal study case to evaluate the efficacy of our proposed framework. Specifically, we will demonstrate the advantage of graph convolutional neural network over traditional order parameters as the state descriptor, and the applicability of deep reinforcement learning for optimal colloidal self-assembly control.

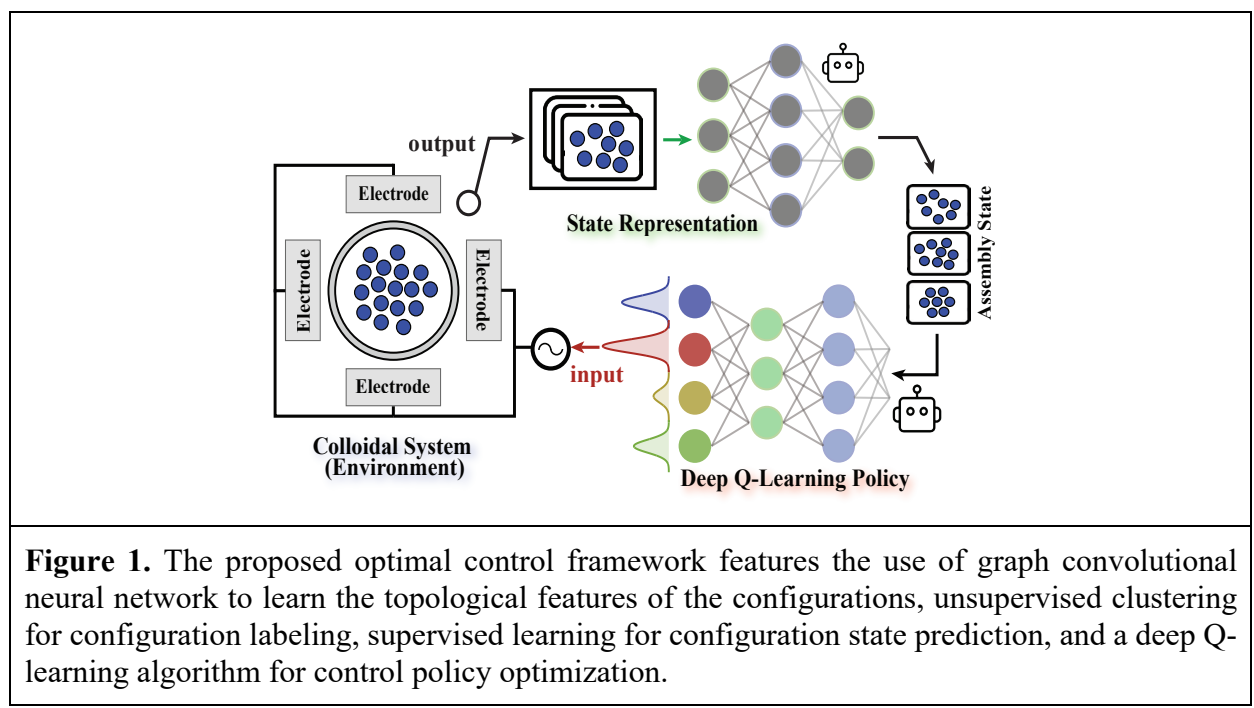

**Figure 1.** The proposed optimal control framework features the use of graph convolutional neural network to learn the topological features of the configurations, unsupervised clustering for configuration labeling, supervised learning for configuration state prediction, and a deep Q-learning algorithm for control policy optimization.

## RESULTS

### Stochastic Dynamics of the Uncontrolled Colloidal System

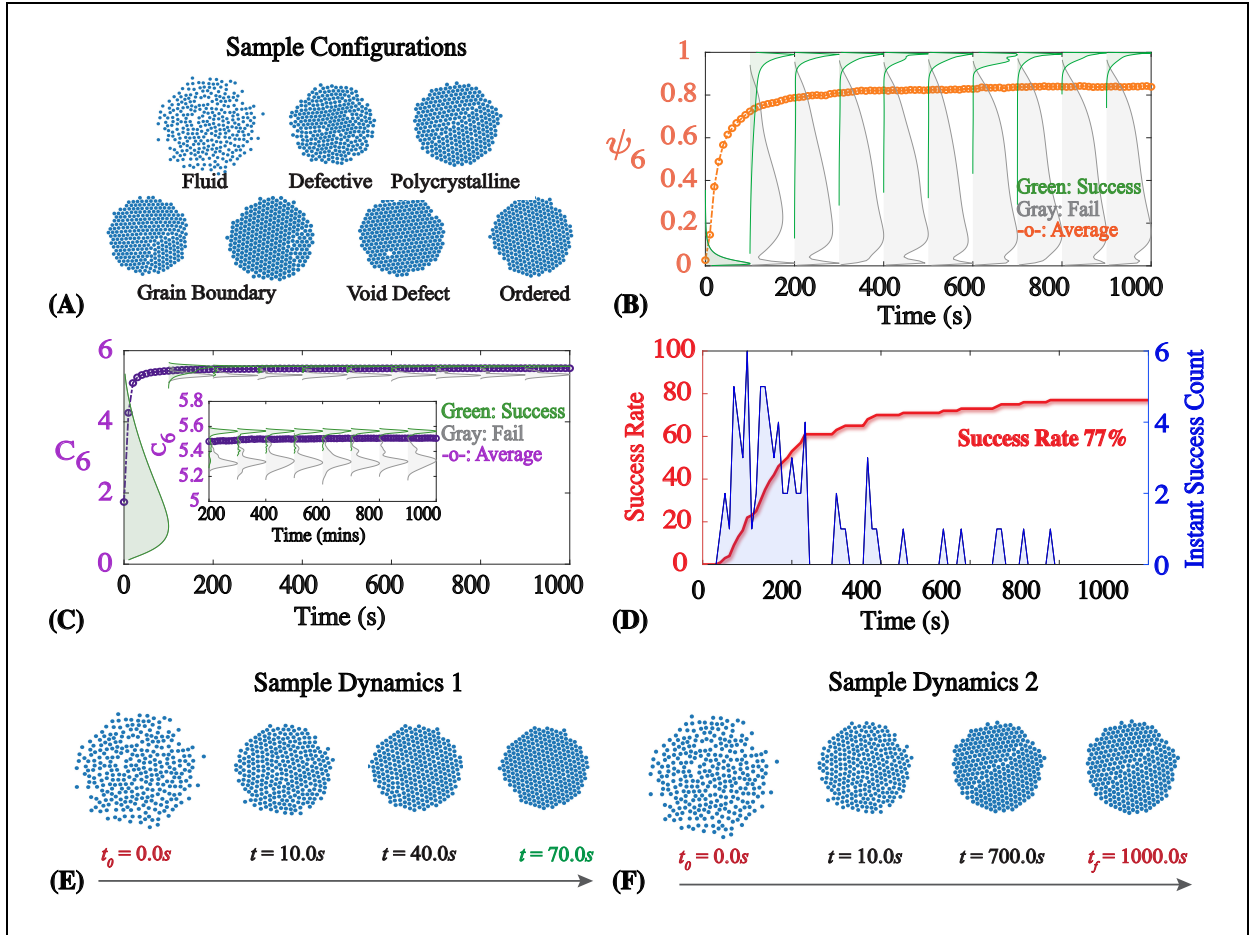

**Figure 2.** Uncontrolled dynamics showing stochasticity of the colloidal system. (A) Representative configurations observed in the system. (B) Distribution (shaded areas) and averaged (orange) evolution of assembly crystallinity in terms of $\psi_6$. (C) Distribution (shaded areas) and averaged (purple) evolution of assembly crystallinity in terms of $C_6$. Note green and gray shaded areas are for success and failed simulations respectively. (D) Cumulative and instant success count over time. (E) Single simulation showing early achievement of the ordered state. (F) Single simulation showing the system stuck in defective states.

We first demonstrate the stochasticity of the system with 100 independently simulated trajectories under the highest voltage level $\lambda = 19.7$, initiated from the similar initial configuration with different seeds for random number generation. For demonstration purpose, we present the dynamics in terms of order parameters $C_6$ and $\psi_6$, and analyze the results in terms of success rate, which quantifies the percentage of simulations that reached the target ordered state at any time during the allotted 1000s of process time, with a criterion of $\psi_6 \geq 0.98$ as previously used in [27, 31].

Starting from a dispersed configuration, the system explored a wide range of representative configurations: fluid-like state, large grain boundary conditions, multi-crystalline states, configuration with void defect, and the ordered configurations (Figure 2A), as the system evolved over the complex free energy landscape under the highest $\lambda$ value [27], and we notice that the averaged order parameters (orange and purple dotted plots in Figure 2B and C) showed a monotonic increase over time, due to the strong compressing force in the system. Simulations that could reach the ordered state showed a rapid jump and stayed at a high value of both $\psi_6$ and $C_6$ early in the simulation (shaded green distribution plots in Figure 2B and C), whereas the ones that never reached the ordered state had a wide distribution of $\psi_6$ values, with noticeably lower $C_6$ value distributions (shaded gray distribution plots in Figure 2B and C), indicating the formation of defective states. Out of the 100 simulations, 77% yielded a crystalline assembly whereas 23% stagnated in defective states and did not evolve into the ordered state within the allotted 1000s of process time, as shown by the cumulative success rate plot in Figure 2D (red). The time needed to achieve the ordered state also exhibited dramatic differences that some simulations only took tens of minutes, whereas some other simulations would take about 900s, according to the distribution of the number of successes over time (shaded blue in Figure 2D), and Figure 2E and F shows two examples of such scenarios.

**DQN-based Optimal Control with Continuous Order Parameters**

To handle the stochasticity in the system, researchers referred to feedback control for a solution, typically using order parameters as the state descriptors. [25, 28, 38] In previous works, the order parameter state space was typically discretized to reduce the dimension of the state for computational efficiency. While applicable, discretizing the originally continuous state space can result in loss of information, compromising the performance of the control. Here we demonstrate the applicability of the DQN algorithm on our system, with a continuous $C_6$-$\psi_6$ state space.

Specifically, we first calculate the order parameter values from 2-dimensional images of the assembly, and then applied DQN to solve for the optimal control action (Figure 3A). To calculate the control policy, we defined the reward function as to maximize the assembly crystallinity, quantified by $\psi_6(t)^2$ as:

$$R(t) = \begin{Bmatrix} \psi_6(t)^2, if\ \psi_6(t)^2 < 0.98 \\ 1 + \psi_6(t)^2, if\ \psi_6(t)^2 \geq 0.98 \end{Bmatrix} \quad \text{Eqn. 1}$$

where an extra reward is given to an action that could lead to the target state, to reinforce the successful strategies. The policy was trained with various and randomly selected initial configurations to facilitate policy convergence and to avoid overfitting to specific initializations, and was solved in PyTorch. The control policy is parameterized by a neural network, which takes the continuous order parameter $C_6$ and $\psi_6$ values as the inputs and outputs a 4-dimensional probability vector, to guide the selection of four control input $\lambda$ values for the next control period $\Delta t = 100s$. At every 100s, the order parameter values will be re-evaluated, and the control action will be updated until the system reaches the target structure or the allotted 1000s of process duration.

Figure 3B presents the optimal control policy projected from the neural network output into the 2-dimensional order parameter state space, where the colors represent different $\lambda$ values. Comparison with the optimal control policy calculated with a model-based dynamic programming using Markov state models developed with discretized $C_6$-$\psi_6$ state space (Figure 3C, referred to as DP-based policy) in [27] for the same system, reveals similarities and significant differences. Both policies featured the highest $\lambda$ input for highly crystalline states (upper right region), and for states with high $C_6$ values, higher $\lambda$ inputs were used as $\psi_6$ increases (top region, from left to right), with the lowest $\lambda$ used in severely defective states (top left region). These observations suggest that in general, a stronger compressing force would be favored as the assembly crystallinity increases, whereas low voltage levels are essential for defects correction. However, in the fluid (lower left) and the transient states (middle range), the DQN-based policy featured a mixed use of different input levels instead of using the highest $\lambda$ value in most of the state space as in the DP-based policy. Several reasons could have contributed to these differences. First, the DP-based policy was calculated based on Markov state models, developed from BD simulations to approximate the BD model with a discretized state space. Due to the unavoidable modeling error, the policy might be optimal for the Markov state model-based system but not for the BD simulations, given the limited samplings for the Markov state model development. Second, as the DQN-based policy presents an approximate dynamic programming approach to solving the continuous state space-based optimal control policy, that due to the infeasibility of sampling all the states, the policy might not have converged to the true global optimality, and the policy here represents one near/sub-optimal solution.

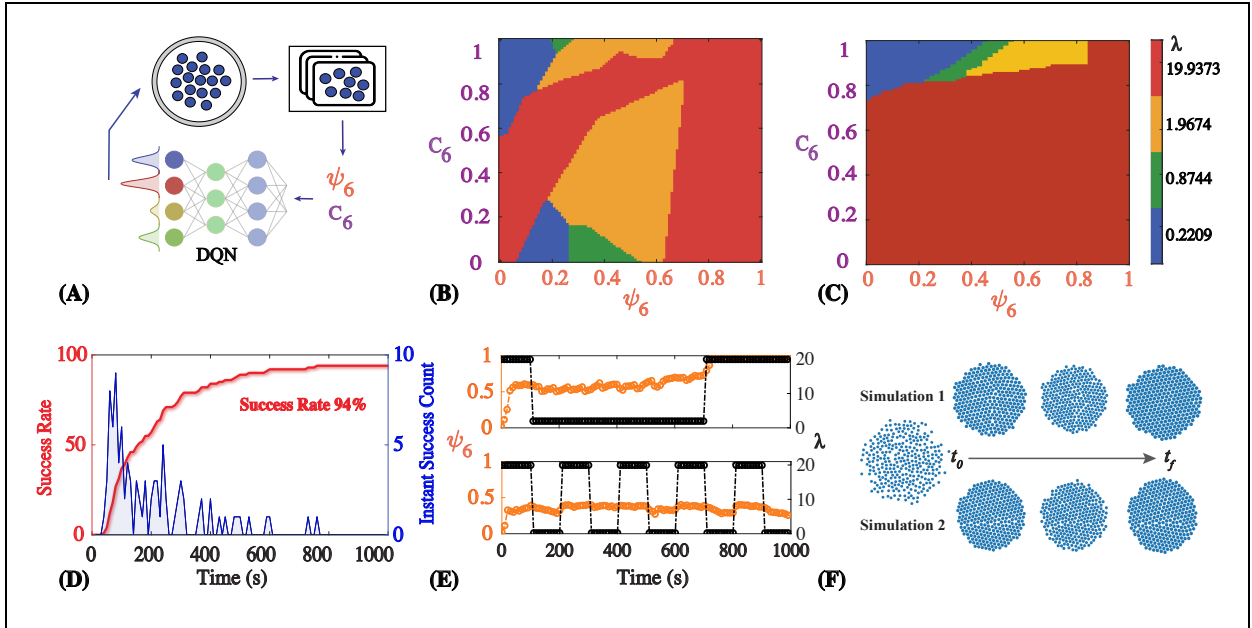

**Figure 3**. DQN-based control policy using continuous order parameters as state descriptors. (A) Schematic of the control setup. (B) 2-D representation of the DQN-based optimal control policy as a lookup table. (C) Reproduced 2-D representation of the dynamic programming-based optimal control policy using discrete order parameters from [27] as a lookup table. $\lambda$ values are denoted in different colors. (D) cumulative and instant success rate of 100 BD simulations controlled with the DQN-based policy showing a success rate of 94%. (E) Two individual controlled BD simulation showing the evolution of $\psi_6$ values and the applied control action $\lambda$ values. (F) Corresponding assembly configuration evolution showing the regulation mechanism.

Despite these differences in the two policies, the results of the 100 independent BD simulation with the DQN-based control policy yielded a 94% success rate in terms of achieving a state with $\psi_6(t) \geq 0.98$, demonstrating a comparably well performance as with the DP-based control policy reported in [27], and showing a significant improvement as compared to simulations without control in Figure 2. Investigating the single controlled trajectories reveals that, 1) relaxation can facilitate the removal of the defects, resonating with previous findings that low voltage can correct defects; and 2) while the policy was effective for most cases, it still failed for certain scenarios that a repetitive use of high and low $\lambda$ values did not remove the defects. Inspecting the assembly configurations further reveals that while order parameters are sufficient in capturing most of the assembly properties, they could misclassify configurations with void defects as perfect crystals, and they could also define a technically crystalline state as a non-perfect state (with $\psi_6 \geq 0.98$), as shown by the representative configurations in Figure 4F. This failure to accurately capturing the configurational information could have been one of the reasons to the failure of the control. Note that, while the control action was updated every 100s, the system dynamics were recorded every 10s for analysis.

**DQN-based Optimal Control with GCN-based State Description**

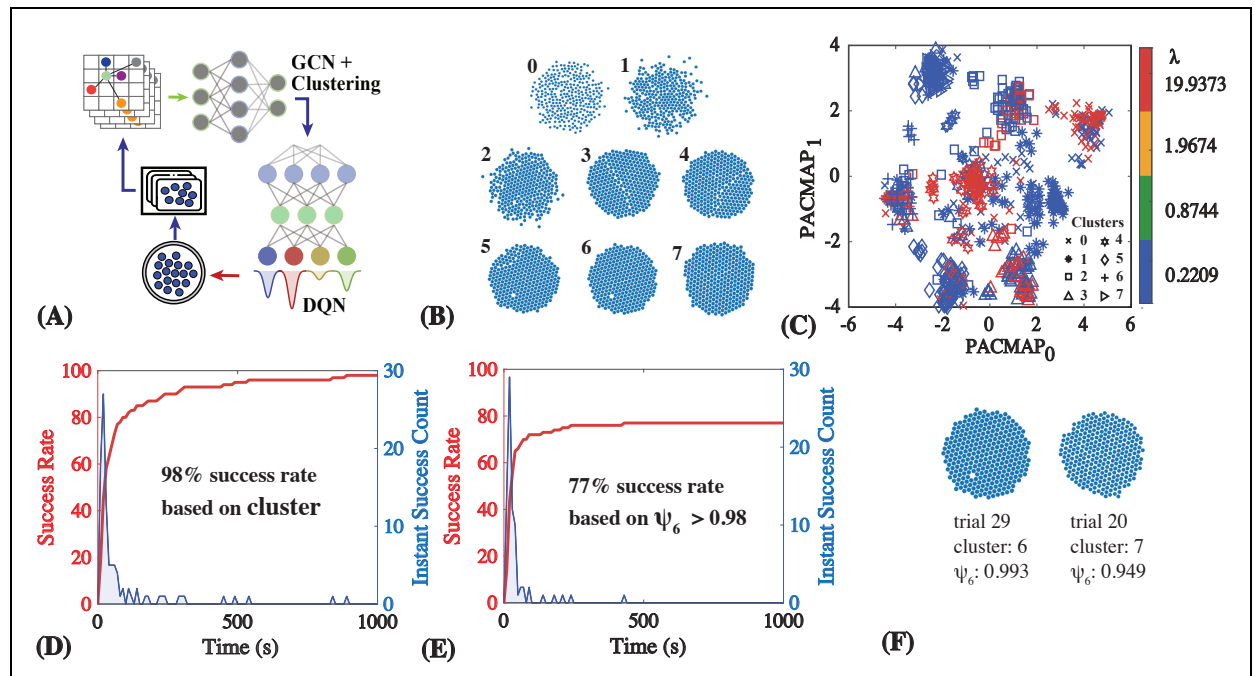

**Figure 4**. DQN-control using GCN-based state representation. (A) Schematic of the control setup featuring GCN and unsupervised clustering for state description and classification. (B) Sample assembly configurations from each of the eight identified clusters/states, with Cluster 7 as the target state. (C) 2-D representation of the 8-D DQN-based optimal control policy in the principal component space. (D) Cumulative and instant success rate of 100 BD simulations controlled with the DQN-based policy showing a success rate of 98%, based on reaching the target Cluster 7. (E) Cumulative and instant success rate of 100 BD simulations controlled with the DQN-based policy showing a success rate of 77%, based on reaching $\psi_6(t) \geq 0.98$. (F) Example configurations showing order parameter can misclassify a void-defect assembly as an ordered state, while classifying an ordered assembly as non-ordered structure.

The controlled results with the DQN-based policy using $C_6$ and $\psi_6$ as the state descriptors confirmed the applicability of DQN to our system, and we then sought to develop and evaluate the performance of the DQN-based control policy, using a graph convolutional neural network (GCN) as the state descriptor.

To construct the GCN, particle positions are extracted from the assembly images and are transformed into a graph, where the nodes represent the particles and the edges represent the connectivity among the particles. To capture the structural features of the assembly, we defined 15 features associated with each particle in the system, resulting in a $N \times F$ dimensional representation of one configuration, where $N = 300$ is the total number of particles in the system, and $F = 15$, is the number of features for each particle. To reduce the dimensionality, we then used an autoencoder to project the output into a 5-dimensional representation, which further serves as the input to an unsupervised clustering step to cluster the configurations into different clusters, i.e. states, following the same procedure as described in [45]. In this work, we used HDBSCAN (Hierarchical Density-Based Spatial Clustering of Applications with Noise) for the unsupervised clustering, we obtained a total of eight discrete clusters as the key states in the system, as shown in Figure 4B. Note that the labeling of the clusters is assigned randomly and bares no physical meaning. To classify a given configuration into one of the eight states, the GCN model then outputs an eight-dimensional probability vector, indicating the probabilities for a given configuration to be classified as each of the eight states. Note that a configuration can be classified in multiple clusters but with different probabilities. After training the GCN model with more than 15000 images, we can then use the GCN prediction to develop the DQN-based control policy. The detailed description of the autoencoder and unsupervised clustering is provided in [45], and the detailed description of the GCN construction is given in the Method section and the SI document.

As the perfect configurations are labeled as Cluster 7, to calculate the control policy, we then defined the reward function as to drive the system to Cluster 7 with a probability higher than 0.95 as the following, to account for the uncertainty in the GCN model:

$$R(t) = \begin{Bmatrix} p(s \text{ in cluster } 7), if\ p(s \text{ in cluster } 7) < 0.95 \\ 1 + p(s \text{ in cluster } 7), if\ p(s \text{ in cluster } 7) \geq 0.95 \end{Bmatrix} \quad \text{Eqn. 2}$$

Same as in Eqn. 1, we added an extra reward to control that has a higher potential to reach the target cluster to reinforce convergence to the successful policies. We then solved for the eight-dimensional optimal control policy with PyTorch, using the Adam optimizer.

Figure 4C shows the 2-dimensional projection of the 8-dimensional control policy after Pairwise Controlled Manifold Approximation (PaCMAP) transformation, for visualization. The markers indicate different clusters, and the colors indicate the corresponding control action $\lambda$ values to use. As compared to the control policy in Figure 3, the GNN-based DQN policy seems to favor more on the highest and the lowest $\lambda$ values, mimicking a bang-bang control strategy. This suggests that control with two inputs instead of four could potentially also yield satisfactory results. Indeed, previous findings on the same system demonstrated that control with only the highest and the lowest $\lambda$ values did achieve near perfect results [31] with an over 95% success rate, using order parameters $C_6$ and $\psi_6$ as the state descriptors, resonating the observations here. Also noticeable in the policy is the scattered use of the lowest $\lambda$ value in all clusters, except for the upper right cluster in the PaCMAP space, which corresponds to the highly crystalline states.

The controlled results in Figure 4D confirmed the efficacy of the policy, with 98 out of 100 simulations reached the target state, i.e. Cluster 7, with an average success time of 93.37s. However, quantifying the success rate in terms of $\psi_6(t) \geq 0.98$ only gave a success rate of 77%, with an average success time of 73.25s. Inspecting the terminal assembly configurations reveals that the GCN-based DQN policy had an actual success rate of 97%, and the low success rate in terms of order parameters was indeed due to the misclassification of the configurations with the order parameters. For example, a configuration with a single void defect can be misclassified as a perfect state as it has a $\psi_6$ value of 0.993, whereas a technically crystalline state can be misclassified as a non-ordered state, as it has a $\psi_6$ value of 0.949. However, the GCN model can actually classify the void-defect configuration into Cluster 6, and the ordered configuration into Cluster 7 (Figure 4F). The quantification of the success rates with respect to the optimization objective and to visual inspection is summarized in Table 1, confirming the efficacy of the proposed GCN-based DQN optimal control framework.

**Table 1**. Statistical comparison across different control strategies, showing the reliability of DQN-based optimal control for achieving the optimization target, and the efficacy of the GCN-based state description in improving the actual success rate of making an ordered hexagonal 2-dimensional colloidal self-assembly. Data are analyzed with 100 independently simulated BD simulations.

| State Representation | Control Strategy | Success Rate | Actual Success Rate |
| --- | --- | --- | --- |
| $C_6, \psi_6$ | $\lambda = 19$ (uncontrolled) | 81% | 77% |
| $C_6, \psi_6$ | DQN | 94% | 88% |
| GCN | DQN | 98% | 97% |

Figure 5 shows three individual simulations to illustrate different control mechanisms with the control policy. The top row shows the control action $\lambda$ and the order parameter $\psi_6$ (as a comparison) over time for each of the three individual simulations, with the corresponding cluster evolutions given in the second row, and the evolutions in terms of configurations given in rows (G)-(I). For some cases, while the system transitions among clusters other than the target cluster, the policy has constantly used the highest $\lambda$ value, leveraging the system stochasticity to resolve the defect (A, D, and G). For void-defect configurations, while the order parameter could not detect, the GCN model correctly captured the formation and the policy intervened the dynamics with low $\lambda$ value for defect removal, eventually leading to an actually ordered configuration (B, E, and H). However, similar to the previous policies, the GCN-based DQN policy also did not yield a 100% success rate as shown in C, F, and I, where the policy intended to leverage the system stochasticity for defect correction but failed within the allotted 1000s of process time. While the convergence issue of reinforcement learning could be one possible reason, another reason could be the controllability of the system, that with the limited control action space (e.g. number of input levels) it could not guarantee a 100% success rate in combating the stochasticity of the system for the ordered state.

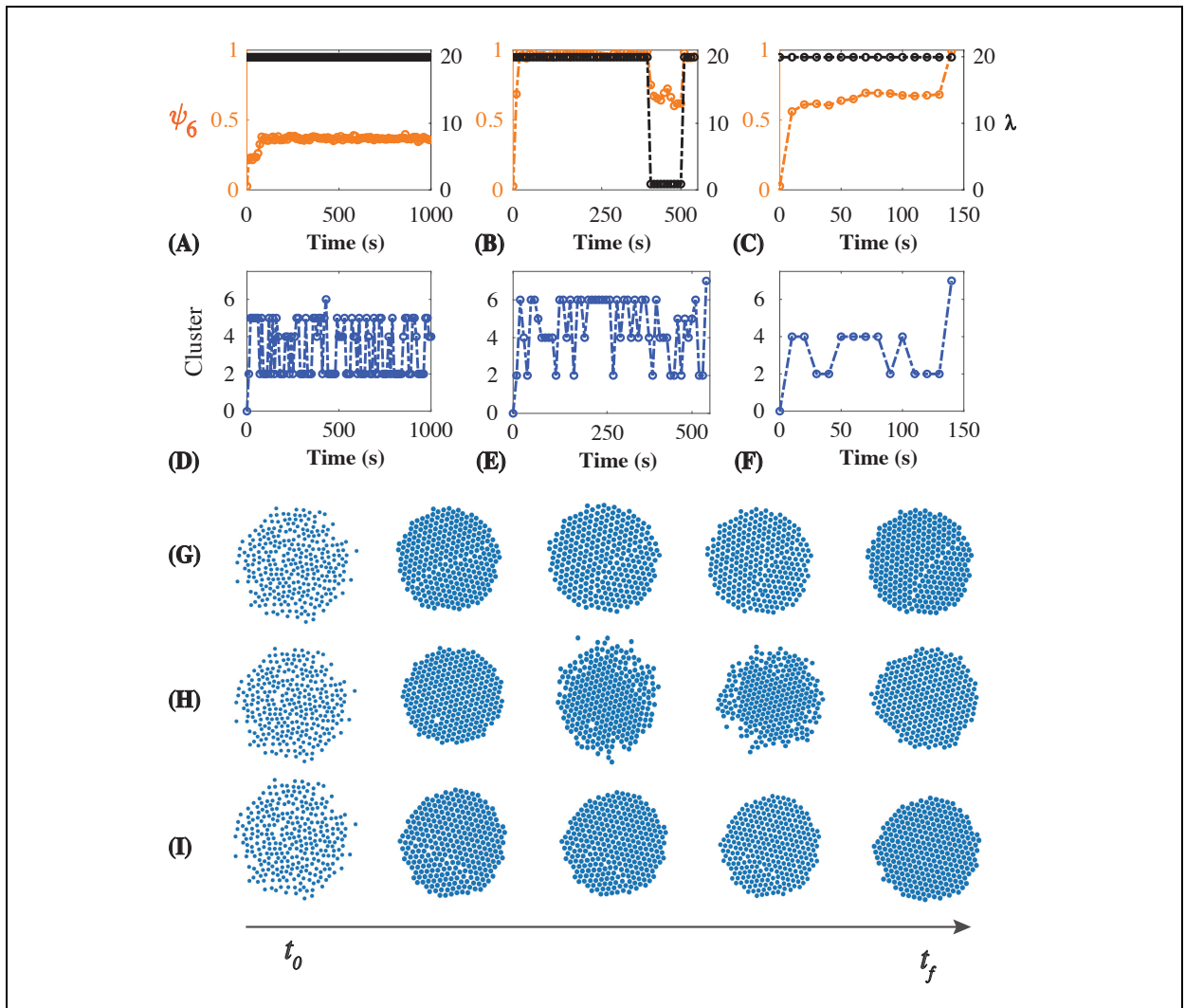

**Figure 5.** Individual controlled simulations showing different dynamics under the control. (A) $\psi_6$ and optimal $\lambda$ values in a simulation where the policy applied the highest $\lambda$ value throughout the process, leveraging the stochasticity of the system for ordering. (B) $\psi_6$ and optimal $\lambda$ value used in a simulation where the policy deployed low $\lambda$ values for void-defect removal. (C) $\psi_6$ and optimal $\lambda$ values used in a simulation where the policy tends to use the system stochasticity for ordering, but failed to achieve the ordered structure. (D)-(F) assembly state evolution in terms of clusters for simulations in (A)-(C) respectively. (G)-(I) assembly evolution in terms of configurations for simulations in (A)-(C) respectively.

## CONCLUSION AND DISCUSSION

Designing an effective control strategy for many-body self-assembly system is challenging, from the state description to system dynamics understanding and to the control policy calculation. In this work, we presented a machine learning-based optimal feedback control framework for the control of a Brownian Dynamics simulation-based quadrupole AC-electric field-mediated 2-dimensional colloidal self-assembly process. Our framework features the use of graph convolutional neural network to extract the configurational features of the assembly for state description, unsupervised clustering for state classification, and deep-Q learning-based reinforcement learning for control policy calculation. Simulation results demonstrated that, control with the DQN-based optimal control policy using the GCN-based state representation can lead to a 97% actual success rate in achieving the target ordered hexagonal assembly, outperforming control that used conventional order parameters as the state representation. This finding confirmed the reliability of using graph convolutional neural network as an alternative state descriptor for colloidal self-assembly, and that DQN-based optimal control is a viable approach to controlling the stochastic process.

As each module of framework is developed with data, we anticipate potential automation of the performance improvement in practical implementation, where part of, or the whole process of the framework can be iteratively repeated for improvement, given the assumption that experimental data collection can be automated via computer programming. While we demonstrated the control performance on achieving a relatively simple target configuration, i.e., the ordered hexagonal structure, in principle, the objective function can be modified to design a policy that targets at other states, permitting application to systems with more complex assembly configurations. Furthermore, as the state representation is achieved by learning the features from particle coordinates or images of the configuration, we also expect the approach to be applicable to systems with large number of particles, and with particles that have different sizes, shapes, and properties.

While effective, we also noticed challenges and directions for further improvement. First, validation of the unsupervised state clustering. Due to the inherent limitation of unsupervised learning, no ground-truth information about the actual number of clusters or the labeling of each configuration is available to assess the accuracy of the clustering, that validating the reliability of the HDBSCAN clustering results remains challenging. Given the importance of state information in feedback control, a small state estimation error could result in unsatisfactory performance. To address this issue, cross-validation of the clustering accuracy with different clustering metrics and approaches should be considered for improvement, and iterative refinement and validation of the clustering and the resulting control policy might be needed for more complicated systems.

Second, during control policy training, we noticed that the last-episode control policy is not necessarily the best policy for the system, due to the stochasticity of the optimization process, suggesting a potential convergence issue with the learning. In our study, we retrieved the reward function trajectory associated with each policy during the training, selected the ones with the highest rewards for evaluation, and identified the best performing one as the optimal policy. While it could be feasible to test a limited number of control policies in certain practical implementations, future efforts should focus on developing more rigorous criteria to facilitate policy selection.

Third, despite its great prospect, reinforcement learning-based control assumes the system dynamics to follow Markovian property, where the future dynamics depends only on current state and control action, regardless of its previous dynamics. While applying reinforcement learning to control colloidal self-assembly has seen successes, for systems with more complex dynamics, such as assembly particles with different shapes and properties in more than one external field, would require validating of the Markovian property of the system dynamics before the implementation of the proposed control framework.

Fourth, the reachability of a given colloidal system needs to be validated. Current colloidal self-assembly for patterned structures normally relies on domain knowledge to first define the settings of the system, such as the properties of the particle (e.g. isotropic vs. anisotropic), the type of driving forces for manipulation (e.g. electric, magnetic or acoustic), and the specific range of the driving force, such as the magnitude of the electric voltage here, and then seek to identify the operation strategies for making the desired configurations. Despite its popularity, it is not guaranteed that such a process can ensure the achievement of the desired configurations, due to the inherent limitations of the system settings, that methods that can evaluate the reachability of the target configuration in a given system should be explored.

Finally, current colloidal self-assembly state descriptors have only focused on the static properties of the assembly, i.e. the assembly configuration, neglecting the kinetics of the system dynamics. For systems with more complex dynamics, such as the ones mentioned above, capturing the static configurational features might not be sufficient for an effective control, that approaches that can also capture the kinetics of the system would be needed. One possible direction could be the complementation with state observers to infer additional system properties besides the static configurational information.

Machine learning has revolutionized a wide range of fields, as studies in machine learning progress and the demand for more complicated colloidal self-assembly structures increases, we expect machine learning to play an increasingly important role in the manufacturing of advanced materials via micron and nano-scaled particle self-assembly.

**METHODS**

**GCN-based state representation and classification**

The GCN-based state representation is achieved with three components: 1) a graph convolutional autoencoder (GCA), which learns low-dimensional embeddings from the graph representation of the colloidal configurations; 2) an unsupervised clustering with HDBSCAN (Hierarchical Density-Based Spatial Clustering of Applications with Noise) to cluster the GCA processed configurations into labeled states; and 3) a supervised learning neural network to classify new configurations as one of the labeled states from the HDBSCAN clustering.

The GCA consists of an encoder and a decoder, designed to learn the compressed representations of graph-structured data through nonlinear graph convolutional operations. The encoder is composed of 2 stacked GCN layers, and maps the high-dimensional input graph structure, which includes the node feature matrix $X$ and the adjacency matrix $A$, to a low-dimensional, dense graph embedding $Z$. In our system, we defined 15 features for each node to describe the graph, including the x and y coordinates of the particle, the six pairwise distance between the six nearest particles, six triangle areas formed with the six nearest particles, and the average of the six triangle areas. The detailed description of the adjacent and feature matrices is given in the SI document.

The propagation rule for the $l$-th GCN layer is defined as:

$$H^{(l+1)} = \sigma\left(\widetilde{D}^{-\frac{1}{2}}\widetilde{A}\widetilde{D}^{-\frac{1}{2}}H^{(l)}W^{(l)}\right)$$

where $H^{(l)}$ denotes the node embeddings at the l-th layer, with $H^{(0)} = X$. $\widetilde{A} = A + I_N$, is the adjacency matrix with self-loops. $\widetilde{D}$ is the corresponding degree matrix of $\widetilde{A}$, $W^{(l)}$ is the trainable weight matrix of the $l$-th layer, and $\sigma$ denotes a nonlinear activation function, and here we used ReLU activation, with two encoding layers, $L = 2$. After the encoding layers, we obtained the final node embeddings $H^{(L)}$. A global average pooling layer is then applied to aggregate all node embeddings into a fixed-size graph-level embedding vector **Z**, which then served as a low-dimensional "fingerprint" of the colloidal configuration for clustering. The decoder is designed to reconstruct the original graph topology from the embedding **Z**, by minimizing the binary cross-entropy loss function:

$$\mathcal{L}_{GCA} = -\frac{1}{N^2} \sum_{i,j=1}^{N} \left[ A_{ij} \log(\hat{A}_{ij}) + (1 - A_{ij}) \log(1 - \hat{A}_{ij}) \right]$$

where $N = 300$, is the number of nodes and $\hat{\boldsymbol{A}}$ is the reconstructed adjacent matrix.

With the graph embeddings **Z** obtained for all the training colloidal configurations from the GCA, we then applied HDBSCAN clustering to classify them into discrete clusters, for the labeling of each of the configurations. This step ultimately partitions the large, unlabeled dataset of colloidal configurations into distinct clusters characterized by unique structural features, providing a foundation for subsequent supervised learning for state identification. In this study, we used the same HDBSCAN setup as detailed in our previous works [45, 46] with a total of ~13,000 training and testing colloidal configurations, following a 70-15-15% training-test-validation split, and obtained a total of 8 clusters, i.e. states.

We then paired the GCA embeddings of each colloidal configuration to its corresponding state labeling, and developed the GCN with supervised learning to classify a given configuration into one of the 8 clusters, by minimizing the cross-entropy loss function:

$$\mathcal{L}_{GCN} = -\sum_{m=1}^{M} \sum_{c=1}^{C} \left[ Y_{mc} \log(\hat{Y}_{mc}) \right]$$

where $M$ is the total number of training colloidal configurations, $C = 8$ is the number of clusters, $\boldsymbol{Y}_{mc} \in \{0,1\}^{M \times C}$ is the ground-truth labeling indicating whether configuration $m$ belongs to cluster $c$ or not, and $\hat{\boldsymbol{Y}}_{mc} \in \{0,1\}^{M \times C}$ is the predicted probability for configuration $m$ to be in cluster $c$, obtained from the softmax activation function:

$$\hat{y} = softmax(H^{(l)} W_{cls})$$

where $\boldsymbol{W}_{cls}$ is the weights of the GCN. The training of all the machine learning models was performed in Python PyTorch, on a Dell Desktop computer with the following specs: 16 Core Processor Intel(R) Core™ i7-10700 CPU @ 2.9GHz, and 1TB SSD.

**Deep Q-learning for Optimal Control Policy**

To solve for the optimal control policy, we implemented a Deep Q-Network (DQN) algorithm using a paired online (actor) network and target network structure in PyTorch. The dimension of the input layer was tailored to match that of the state representation. For example, for the order parameter-based state representation, the network received two scalar inputs ($C_6$ and $\psi_6$), whereas for the GCN-based representation, the network received a vector that contains the eight cluster probabilities obtained from the GCN prediction. For both scenarios, the network consisted of two fully connected hidden layers with 256 and 128 neurons, respectively, each followed by a ReLU activation. The output layer was a linear layer with four nodes corresponding to the four discrete control actions.

Each control policy was trained for at least 5000 episodes, with one episode corresponding to a 1000s Brownian Dynamics simulation, starting from 10 randomly selected initial configurations, covering fluid, defective and ordered configurations. The target network remained fixed during each update cycle and was synchronized with the actor network every 10 episodes, to improve training stability by preventing harmful feedback loops between Q-value prediction and target estimation. The Smooth L1 (Huber) loss computed between the predicted Q-values and the target Q-values was used as the loss function used to update the actor network.

The Bellman optimality equation for the Q-function is given as:

$$Q^*(s,a) = E\left[r + \gamma \max_{a'} Q^*(s',a') | s,a\right]$$

where $r$ is the reward received after taking a control action $a$, customized to fit the state representation as specified above. $s'$ is the next time step state, $\gamma = 0.99$ is the discount factor, and $a'$ is the next time step control action. In the DQN, we trained a neural network $Q_\theta(s,a)$ parameterized by $\theta$, to approximate $Q^*(s,a)$, by minimizing the mean squared error between the current Q-value estimate and a target value derived from the Bellman equation.

$$\mathcal{L}_Q = E_{(s,a,r,s') \sim D}[(y - Q_\theta(s,a))^2]$$

With the target value $y$ defined as: $y = r + \gamma \max_{a'} Q_{\theta^-}(s',a')$, where $\theta^-$ denotes the parameters of the target network that is periodically updated to stabilize the training, and $D$ is the replay buffer that stores $(s,a,r,s')$.

Each training episode was terminated upon either reaching the target state or exceeding 100 BD simulation steps, with one step corresponding to 10s of process time. Action selection followed an $\varepsilon$-greedy strategy as the following, with $\varepsilon$ initialized at 0.99 and decaying exponentially with a decay constant of $10^{-4}$ to ensure adequate exploration.

$$a_t = \begin{cases} random\ action & with\ probability\ \varepsilon, \\ argmax_a Q_\theta(s_t,a) & with\ probability\ 1-\varepsilon. \end{cases}$$

After each episode, the actor network was then updated using a minibatch of 100 transitions sampled randomly from the replay buffer $D$.

All the DQN-based control policies were solved with PyTorch *Adam* optimizer on a Mac Pro M2 Ultra, with the following specs: 24-Core CPU, 60 Core GPU, 192GB RAM, and 8 TB SSD. Training the control policy took about 1.5 weeks, which this was mainly due to solving the BD simulations.

**Data Availability**

The control policy and the associated controlled BD simulation data that support the findings of this study are available at:

https://github.com/xtang38/Lizano_Ma_et_al_GCN_based_DQN_control.

**Code Availability**

The essential codes on GCN, autoencoder, clustering, and the DQN setting are accessible at: https://github.com/xtang38/Lizano_Ma_et_al_GCN_based_DQN_control.

## AUTHOR INFORMATION

### Corresponding Author

Xun Tang: xuntang@lsu.edu

### Author Contributions

X.T. conceived the concept, supervised the work, and analyzed the results. A.L. and F.M. performed the study and analyzed the results. W.S. and W.T. analyzed the results. X.T., A.L. F.M. wrote the original draft, all authors contributed to the editing of the manuscript and have given approval to the final version of the manuscript.

‡These authors contributed equally.

### Funding Sources

A.L. and X.T. are supported by NSF grant #2218077; W.T. is supported by NSF grant #2414369.

## ACKNOWLEDGMENT

We thank Dr. Michael Bevan in the Chemical & Biomolecular Engineering at Johns Hopkins University for sharing the BD model for this study.

## ABBREVIATIONS

BD: Brownian Dynamics; GCN: Graph Convolutional Neural Network; GCA: Graph Convolutional Autoencoder; DQN: Deep Q-Learning; DP: Dynamic Programming. PaCMAP: Pairwise Controlled Manifold Approximation; HDBSCAN: Hierarchical Density-Based Spatial Clustering of Applications with Noise.